\documentclass[]{spie}  
\usepackage{graphicx}
\usepackage{amsmath,amssymb}  
\usepackage{mathtools} 
\usepackage{bm}  
\usepackage{enumerate}
\usepackage{xspace}

\newcommand{\qed}{\hfill \ensuremath{\Box}}
\newcommand{\vMF}{von\phantom{-}Mises-Fisher\xspace}

\title{Classification and construction of closed-form kernels for signal representation on the 2-sphere} 

\author{Rodney A. Kennedy\supit{a}, Parastoo Sadeghi\supit{a}, Zubair Khalid\supit{a} and Jason D. McEwen\supit{b}
\skiplinehalf
\supit{a}Research School of Engineering, The Australian National University, Canberra, Australia \\
\supit{b}Department of Physics and Astronomy, University College London, London WC1E~6BT, and
Mullard Space Science Laboratory (MSSL), University College London, Surrey RH5~6NT, U.K.
}

\authorinfo{Email: rodney.kennedy@anu.edu.au}

\newcommand{\ltwo}{\ell{}^{\,2}}
\newcommand{\untsph}{\mathbb{S}^{2}} 
\newcommand{\lsph}{L^2(\untsph)}
\newcommand{\boldhat}[1]{\widehat{\bm{#1}}}
\newcommand{\unit}[1]{\boldhat{#1}}
\newcommand{\conj}[1]{\overline{#1}} 
\newcommand{\dfn}{\coloneqq}
\newcommand{\intsph}{\int_{\mathrlap{\untsph}}\hspace{2mm}} 
\newcommand{\op}[1]{\mathcal{#1}}
\newcommand{\figref}[1]{Fig.\,\ref{#1}}
\newcommand{\Hspc}{\mathcal{H}} 
\newcommand{\Kspc}{\Hspc_{K}^{\vphantom{p}}(\untsph)} 
\newcommand{\tendsto}{\rightarrow}
\newcommand{\longtendsto}{\longrightarrow}

\DeclarePairedDelimiterX\abs[1]{\lvert}{\rvert}{#1}
\DeclarePairedDelimiterX\brak[1]{\lbrace}{\rbrace}{#1}
\DeclarePairedDelimiterX\cardinality[1]{\lvert}{\rvert}{#1}
\DeclarePairedDelimiterX\innerp[2]{\langle}{\rangle}{#1,#2}
\DeclarePairedDelimiterX\norm[1]{\lVert}{\rVert}{#1}
\DeclarePairedDelimiterX\parn[1]{(}{)}{#1}
\DeclarePairedDelimiterX\sequ[1]{\lbrace}{\rbrace}{#1}
\DeclarePairedDelimiterX\set[1]{\lbrace}{\rbrace}{#1}
\DeclarePairedDelimiterX\sqrb[1]{[}{]}{#1}
\DeclarePairedDelimiterX\coeff[1]{(}{)}{#1}

\begin{document} 
  \maketitle 

\begin{abstract}
This paper considers the construction of Reproducing Kernel Hilbert Spaces~(RKHS) on the sphere as an alternative to the conventional Hilbert space using the inner product that yields the $\lsph$ function space of finite energy signals. In comparison with wavelet representations, which have multi-resolution properties on $\lsph$, the representations that arise from the RKHS approach, which uses different inner products, have an overall smoothness constraint, which may offer advantages and simplifications in certain contexts. The key contribution of this paper is to construct classes of closed-form kernels, such as one based on the \vMF distribution, which permits efficient inner product computation using kernel evaluations.  Three classes of RKHS are defined: isotropic kernels and non-isotropic kernels both with spherical harmonic eigenfunctions, and general anisotropic kernels.
\end{abstract}



\section{Introduction}
\label{sec:intro}  

The main purpose of the present paper is to provide an accessible framework where closed-form kernels can be constructed and interpreted so as to define various Reproducing Kernel Hilbert Spaces~(RKHS) on the sphere.  By closed-form we mean that the kernel can be expressed in terms of elementary or special functions\cite{Lebedev:1972} and only involve a finite number of such functions (and preferably just one) --- the well-known reason for seeking such closed-form kernels is that inner product evaluations in the RKHS can be performed in terms of kernel evaluations.  This leads to a classification of kernels into three classes: 1) isotropic kernels with spherical harmonics as the eigenfunctions; 2) anisotropic kernels also with spherical harmonics as the eigenfunctions; and finally 3) the most general anisotropic case.  In the most specific class we merge known results with the development of some novel closed-form kernels and leave open the development of yet further kernels.

Analysis of signals defined on the 2-sphere, $\untsph$, is dominated by techniques using the spherical harmonic transform\cite{Kennedy-book:2013}.  The spherical harmonic transform is the natural generalization of Fourier series using complex exponentials for signals defined on a bounded closed interval on the real line.  With a standard definition of inner product, the spherical harmonics are a complete and orthonormal sequence of functions defined on the 2-sphere and are closely tied with the study of the class of finite-energy signals generally represented by $\lsph$.  This space is not without some problems as the member vectors need not be strictly functions but equivalence classes of functions which are almost everywhere equivalent.  Or possibly the vectors are ones that, because of their high variations, would never appear in practice.  This means that there are functions in $\lsph$ which do not represent meaningful signals.  Further, in computation aspects the inner product computations can be formidable even in the spherical harmonic domain where FFT methods can be used.  So there is interest in considering a different space of smooth functions with reduced complexity to represent signals and retain some simplicity in the computation of the inner product.

The theory of reproducing kernel Hilbert spaces was developed by Aronszajn\cite{Aronszajn:1950}.  They are a Hilbert space where the vectors are functions and a special inner product is defined which yields a number of interesting and useful properties.  Among these properties we note that: the vectors are continuous functions; a kernel function can be used to represent these functions; and the evaluation of the inner product between two vectors can be expressed in terms of an evaluation of the kernel function which avoids an implicit computation in an infinite dimensional space.  This computational aspect does however require that the kernel itself is easy to compute which translates to the property that the kernel should be ``closed-form'' in the sense that it can be expressed in terms of a elementary or special functions\cite{Lebedev:1972} and only involve a finite number of such functions.  Often the space is designed first by specifying a closed-form candidate kernel function but such a function needs to satisfy a positive definite condition.  Conversely working from positive definite conditions generally does not lead to a closed-form kernel.  It is this context that we seek closed-form kernels for signals defined on the 2-sphere.

In this paper we lay down the conditions that a kernel needs to satisfy to lead to a RKHS on the 2-sphere.  We classify the three classes of kernels that may arise.  Then a special case of isotropic kernels is considered and their general properties derived.  For this special case we find a set of closed-form kernels with various properties.

\begin{figure}
\centering
	\includegraphics[width=0.5\columnwidth]{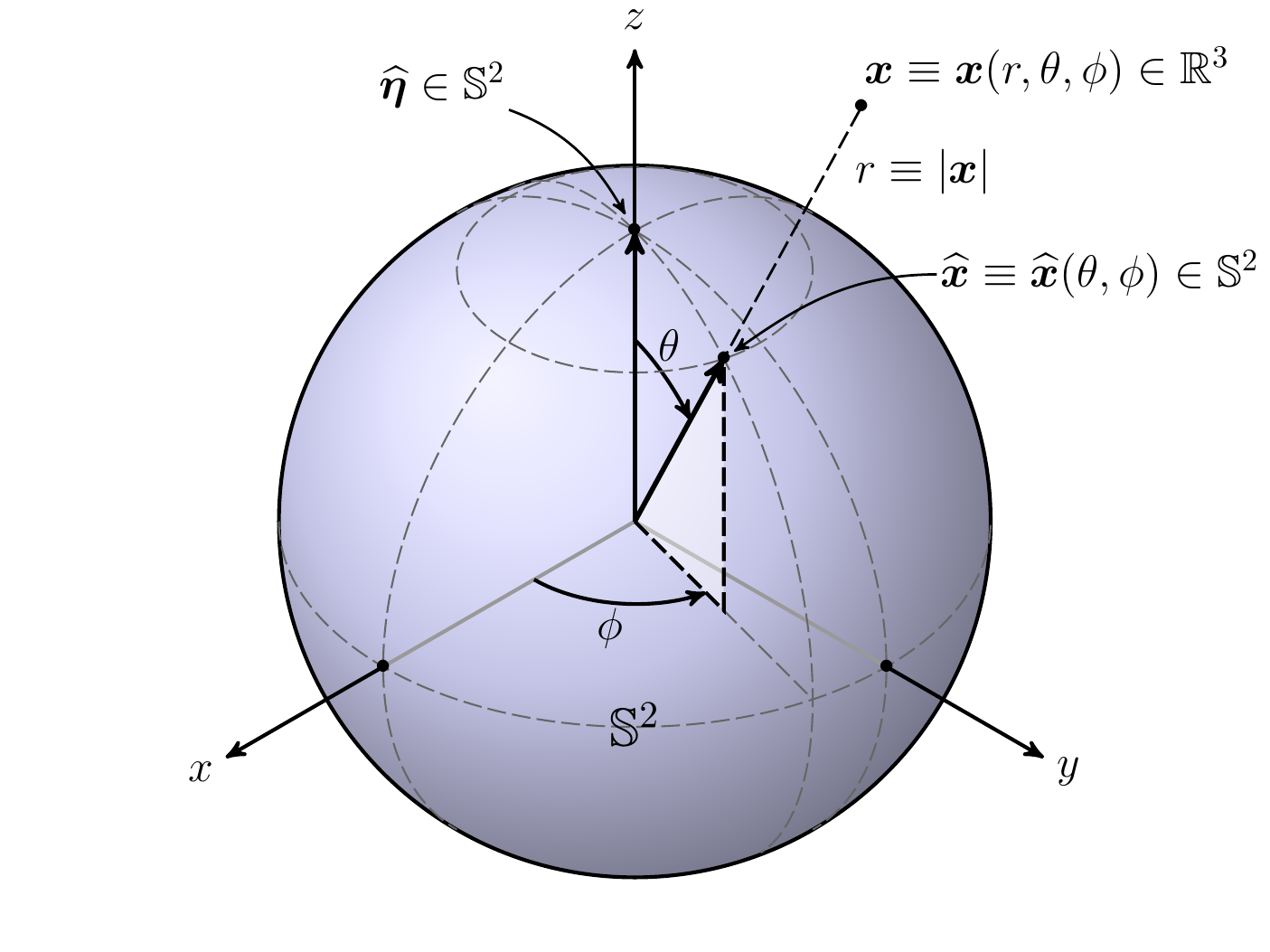}\\
	\caption{Coordinate system showing a point in 3D space $\bm{x}\in\mathbb{R}^{3}$, a unit vector $\unit{x}$, which lies on the 2-sphere, $\untsph$, north pole $\unit{\eta}\in\untsph$, the spherical polar coordinate system with $\theta\in[0,\pi]$ the co-latitude, $\phi\in[0,2\pi)$ the longitude and $r\equiv\abs{\bm{x}}$ the Euclidean distance.}\label{fig:s2-coords}
\end{figure}

\section{Construction and Classification of RKHS on 2-Sphere}

\subsection{RKHS Construction}
\label{sec:rkhscon}

We begin with the complex Hilbert space finite energy functions on the 2-sphere, $\lsph$, which is equipped with the standard inner product for functions $f,g$
\begin{equation}
\label{eqn:L2-ip}
	\innerp{f}{g}\dfn \intsph f(\unit{x}) \conj{g(\unit{x})}\,ds(\unit{x}),
\end{equation}
where, see \figref{fig:s2-coords}, $\bm{x}\in\mathbb{R}^{3}$, $\unit{x}\dfn (\cos\phi\sin\theta,\sin\phi\sin\theta,\cos\theta)'\in\untsph\subset\mathbb{R}^{3}$, $\untsph\dfn\set{\bm{x}\in\mathbb{R}^{3}\colon \abs{\bm{x}}=1}$, $\theta$ is the co-latitude, $\unit{\eta}\in\untsph$ is the north pole, $\phi$ is the longitude, and $ds(\unit{x})=\sin\theta\,d\theta\,d\phi$ is the uniform surface measure such that $\intsph\, ds(\unit{x}) = 4\pi$.  Finite energy functions are those that satisfy the bounded induced norm condition
\[
	f\in\lsph \iff \norm{f}\dfn\innerp{f}{f}{}^{1/2}<\infty.
\]
In this work we require the spherical harmonics.  They are defined through
\[
	Y_{\ell}^{m}(\theta,\phi) \dfn
		\sqrt{\frac{2{\ell}+1}{4\pi}\frac{({\ell}-m)!}{({\ell}+m)!}}
		P_{\ell}^{m}(\cos\theta)e^{im\phi}, \text{\quad equivalently denoted as\quad} Y_{\ell}^{m}(\unit{x}),
\]
where $\ell\in\set{0,1,\dotsc}$ is the degree, $m\in\set{-\ell,-\ell+1,\dotsc,\ell}$ is the order, the associated Legendre functions are
\[
	P_{\ell}^{m}(z) \dfn \frac{(-1)^m}{2^{\ell}{\ell}!} (1-z^{2})^{m/2}
		\frac{d^{{\ell}+m}}{dz^{{\ell}+m}}
		(z^{2}-1)^{\ell},\quad
		m\in\set{0,1,\dotsc,\ell}.
\]
and satisfy
\[
	P_{\ell}^{-m}(z) = (-1)^{m} \frac{({\ell}-m)!}{({\ell}+m)!} P_{\ell}^{m}(z),
		\quad m\in\set{0,1,\dotsc,\ell},
\]
which enables the determination of the spherical harmonics for $m\in\set{-1,-2,\dotsc,-\ell}$.  Finally, completeness of the spherical harmonics means
\[
	f(\unit{x}) = \sum_{\ell,m}
		\innerp{f}{Y_{\ell}^{m}} Y_{\ell}^{m}(\unit{x}),\quad\forall\,f\in\lsph \text{\quad where }
		\sum_{\ell,m} \dfn \sum_{\ell=0}^{\infty} \sum_{m=-\ell}^{\ell}.
\]
For a construction of a specific class of RKHS's on the 2-sphere based on the spherical harmonics, which are orthonormal in $\lsph$, we present an adaptation of standard RKHS theory\cite{Cucker:2002}.

\vspace{3mm}
\begin{theorem}
\label{thm:shrkhs}
Define a square summable sequence of positive real weights
\begin{equation}
\label{eqn:lambda-seq}
	\sequ[\big]{\lambda_{\ell}^{m}\colon \lambda_{\ell}^{m}\in\mathbb{R},
		\lambda_{\ell}^{m}>0}_{\ell,m}\in\ltwo,
\end{equation}
and corresponding sequence of spherical harmonics
\begin{equation}
\label{eqn:spharm-seq}
	\sequ[\big]{Y_{\ell}^{m}(\unit{x})}_{\ell,m}
\end{equation}
with matching degrees $0\leq\ell<\infty$ and orders $-\ell\leq m\leq \ell$.

Then the set of functions on the 2-sphere satisfying the following induced norm condition forms a Hilbert space, $\Kspc$,
\[
	f\in\Kspc \iff \norm{f}_{\Kspc}\dfn\innerp{f}{f}{}_{\Kspc}^{1/2}<\infty
\]
with inner product
\begin{equation}
\label{eqn:yrkhs-ip}
	\innerp{f}{g}_{\Kspc} \dfn \sum_{\ell,m}
		\frac{\innerp{f}{Y_{\ell}^{m}}\innerp{Y_{\ell}^{m}}{g}}{\lambda_{\ell}^{m}}.
\end{equation}
Further, if we define the positive definite kernel
\begin{equation}
\label{eqn:yrkhs-kernel}
	K(\unit{x},\unit{y}) \dfn \sum_{\ell,m}
		\lambda_{\ell}^{m} \, Y_{\ell}^{m}(\unit{x}) \conj{Y_{\ell}^{m}(\unit{y})}
\end{equation}
then we have the properties:
\begin{enumerate}[\bfseries~(P1)]
\item\label{itm:p1} Hermitian symmetry property:
\begin{equation}
\label{eqn:p1}
	K(\unit{x},\unit{y}) = \conj{K(\unit{y},\unit{x})},
\end{equation}
\item Reproducing kernel property:
\begin{equation}
\label{eqn:p2}
	\innerp[\big]{f(\cdot)}{K(\cdot,\unit{y})}_{\Kspc} = f(\unit{y}),\quad\forall\,f\in\Kspc,
\end{equation}
\item\label{itm:p3} Kernel property:
\begin{equation}
\label{eqn:p3}
	K(\cdot,\unit{y})\in\Kspc,\quad \forall\,\unit{y}\in\untsph
\end{equation}

\item\label{itm:p4} Eigen-function property:
\begin{equation}
\label{eqn:p4}
	\parn[\big]{\op{L}_{K}Y_{\ell}^{m}}(\unit{x}) = \lambda_{\ell}^{m}\,Y_{\ell}^{m}(\unit{x}),\quad
		\forall\,\ell,m
\end{equation}
where
\begin{equation}
\label{eqn:integ-op}
	\parn[\big]{\op{L}_{K}f}(\unit{x}) \dfn \intsph K(\unit{x},\unit{y})\,f(\unit{y})\,ds(\unit{y}).
\end{equation}

\item Strictly positive operator property:
\begin{equation}
\label{eqn:p5}
	\innerp{\op{L}_{K}f}{f}>0,\quad\forall\,f\in\Kspc \text{ satisfying } f\neq o,
\end{equation}
where $o\in\Kspc$ is the zero vector.
\end{enumerate}
Moreover, $\Kspc$ consists only of continuous functions.
\end{theorem}

\vspace{3mm}

\noindent{\bf Proof:}
Property (P\ref{itm:p1}) follows directly from \eqref{eqn:yrkhs-kernel} and the property that $\lambda_{\ell}^{m}\in\mathbb{R}$.  For property (P\ref{itm:p3}) , one has $\smash{\norm{K(\cdot,\unit{y})}_{\Kspc}=K(\unit{y},\unit{y})<\infty}$ by continuity and boundedness.  Property (P\ref{itm:p4}) follows directly from \eqref{eqn:yrkhs-kernel} substituted into \eqref{eqn:integ-op} and letting $f=Y_{\ell}^{m}$.  The remaining parts of the proof can be found in the literature\cite{Cucker:2002,Kennedy-book:2013}.\qed




\subsection{Isomorphism and Integral Operator Properties}

It is important to emphasis that the only difference between a conventional Hilbert space of functions such as $\lsph$ and a RKHS of functions defined on the same domain such as given in $\Kspc$ is in the representation of the vectors and the definition of the inner product.  All Hilbert spaces of the same dimension over the same scalar field are isomorphic\cite{Kennedy-book:2013}.  The vectors in one Hilbert space can be mapped to vectors in the other Hilbert space in a way that evaluations of inner products (and norms) are preserved, and the Fourier coefficients are equal.  The isomorphism between the $\lsph$ and $\Kspc$ is given by
\begin{align}
	\label{eqn:isomap1}
	\op{L}_{K}^{-1/2}\colon \Kspc &\longtendsto \lsph \\
	f &\longmapsto h \nonumber
\shortintertext{and}
	\label{eqn:isomap2}
	\op{L}_{K}^{1/2}\colon \lsph &\longtendsto \Kspc \\
	h &\longmapsto f \nonumber
\end{align}
where $\op{L}_{K}^{1/2}$ is the strictly positive square root of the operator $\op{L}_{K}$ in \eqref{eqn:integ-op} satisfying $\op{L}_{K}=\op{L}_{K}^{1/2}\circ\op{L}_{K}^{1/2}$, and $\op{L}_{K}^{-1/2}$ is its inverse (here $\circ$ denotes operator composition).  These mappings are straightforward once one recognizes that
\[
	\parn[\big]{\op{L}_{K}^{p}f}(\unit{x}) = \sum_{\ell,m} (\lambda_{\ell}^{m})^{p}
		\innerp{f}{Y_{\ell}^{m}}\,Y_{\ell}^{m},
\]
for any real power $p\in\mathbb{R}$, noting that $\lambda_{\ell}^{m}>0$ (positive real).  Note that $\op{L}_{K}$ is compact, but $\op{L}_{K}^{1/2}$ need not be.  We can expect that $\op{L}_{K}^{-1/2}$ is not bounded.

\subsection{Vector Expansion using Orthonormal Sequence}

We now are interested in how to represent the vectors in the RKHS $\Kspc$.  The spherical harmonics can be used in a Fourier synthesis expansion to represent vectors in $\lsph$ since they are complete and orthonormal.  They have many ideal properties that have made them a compelling representation and have dominated applications and theory\cite{Colton:2013}.  With $\Kspc$ it might be regarded as a death-nell for this new Hilbert space if we have to abandon spherical harmonics for something less workable.  Fortunately that is not the case.

For $\Kspc$ the sequence
\begin{equation}
\label{eqn:Kspcons}
	\varphi_{\ell}^{m}(\unit{x}) \dfn \sqrt{\lambda_{\ell}^{m}}\,Y_{\ell}^{m}(\unit{x})
\end{equation}
is orthonormal under inner product \eqref{eqn:yrkhs-ip} and can be shown to be complete by manipulating the inner product expression \eqref{eqn:yrkhs-ip}.  Therefore we have
\begin{align*}
	f(\unit{x}) &= \sum_{\ell,m} \innerp{f}{\varphi_{\ell}^{m}}_{\Kspc}\varphi_{\ell}^{m}(\unit{x})
		\shortintertext{and}
	f\in\Kspc &\iff \sequ[\big]{\innerp{f}{\varphi_{\ell}^{m}}_{\Kspc}}_{\ell,m}\in\ltwo.
\end{align*}
Hence, we see that the spherical harmonics are preserved in shape, are orthogonal but are not normalized in $\Kspc$.  It is because the spherical harmonics are attenuated for higher degrees $\ell$ (noting that condition \eqref{eqn:lambda-seq} implies $\lambda_{\ell}^{m}\tendsto0$ as $\ell\tendsto\infty$) relative to their mapped isomorphic counterparts in $\lsph$, \eqref{eqn:isomap1}, that the functions in $\Kspc$ are smoother, low-pass and indeed continuous.

Inner products, \eqref{eqn:yrkhs-ip}, in $\Kspc$, can be alternatively evaluated in the RKHS Fourier domain using the orthonormal sequence in $\Kspc$, \eqref{eqn:Kspcons},
\[
	\innerp{f}{g}_{\Kspc} \dfn
		\sum_{\ell,m} \innerp{f}{\varphi_{\ell}^{m}}_{\Kspc} \conj{\innerp{g}{\varphi_{\ell}^{m}}}_{\Kspc}.
\]
Note that inner product \eqref{eqn:yrkhs-ip} is defined in terms of the $\lsph$ Fourier domain.

\subsection{Vector Expansion using Kernel}

An RKHS with kernel $K(\unit{x},\unit{y})$ offers a second way to represent vectors because of the two properties \eqref{eqn:p2} and \eqref{eqn:p3}.  First we rewrite \eqref{eqn:p3} as follows
\begin{equation}
\begin{split}
	K_{\unit{y}}(\unit{x}) &\dfn K(\unit{x},\unit{y}), \quad \forall\,\unit{x},\unit{y}\in\untsph \\
	K_{\unit{y}} &\in \Kspc, \quad \forall\,\unit{y}\in\untsph
\end{split}
\end{equation}
Now suppose we have a finite number $P$ of points on the 2-sphere, that is, $\unit{y}_{p}\in\untsph$, and corresponding samples $\alpha_{p}\in\mathbb{C}$, for $p=1,2,\dotsc,P$, such that we can define
\begin{equation}
\label{eqn:fexpan}
	f(\cdot) \dfn \sum_{p=1}^{P} \alpha_{p}\,K_{\unit{y}_{p}}(\cdot)
		 = \sum_{p=1}^{P} \alpha_{p}\,K(\cdot,\unit{y}_{p}) ,\quad f\in \Kspc.
\end{equation}
Similarly, suppose we have $Q$ samples $\beta_{q}\in\mathbb{C}$ located at $\unit{x}_{q}\in\untsph$, for $q=1,2,\dotsc,Q$, such that we can define
\begin{equation*}
	g(\cdot) \dfn \sum_{q=1}^{Q} \beta_{q}\,K_{\unit{x}_{q}}(\cdot)
		 = \sum_{q=1}^{Q} \beta_{q}\,K(\cdot,\unit{x}_{q}) \in \Kspc.
\end{equation*}
Then the reproducing property \eqref{eqn:p2} leads to
\begin{equation}
\label{eqn:innerp-ker}
	\innerp[\big]{f(\cdot)}{g(\cdot)}_{\Kspc} = \sum_{p=1}^{P} \sum_{q=1}^{Q}
		\alpha_{p}\conj{\beta_{q}}\,K(\unit{x}_{q},\unit{y}_{p}).
\end{equation}

Because the inner product when using the kernel expansion for a vector requires the evaluation of the kernel then this motivates having a kernel that is expressed in terms of elementary or special functions.\cite{Lebedev:1972}

\subsection{RKHS Classification}

Not every positive definite kernel $K(\unit{x},\unit{y})$ on the sphere can be put in the form \eqref{eqn:yrkhs-kernel} because we have restricted attention to using spherical harmonics as our orthonormal functions.  However, the theorem can be generalized by using any orthonormal sequence $\sequ{\varphi_{n}(\unit{x})}_{n\in\mathbb{Z}}$ which is complete in $\lsph$, in place of the sequence of spherical harmonics given in \eqref{eqn:spharm-seq}; and indexing with positive real $\sequ{\lambda_{n}}{}_{n\in\mathbb{Z}}$ sequence replacing \eqref{eqn:lambda-seq}.  This generalization is possible because the operator $\op{L}_{K}$ in \eqref{eqn:integ-op} is compact and self-adjoint and by the spectral theorem has a countable number of eigenfunctions that are orthogonal and complete in $\lsph$ which can be taken as $\sequ{\varphi_{n}(\unit{x})}_{n\in\mathbb{Z}}$.  That is, operator $\op{L}_{K}$ uniquely determines the eigenfunctions and eigenvalues, and vice versa.  In summary, we have three classes of RKHS which depend of the type of eigenfunctions and an isotropic property which will be fleshed out in a later section:
\begin{enumerate}[\bfseries {Class} 1.]
\item
Isotropic kernels where the kernel $K(\unit{x},\unit{y})$ reduces to a function of the 3D dot product $\unit{x}\cdot\unit{y}$, and the eigenfunctions are the spherical harmonics, and the eigenvalues repeat in a way, $\lambda_{\ell}^{m}=\lambda_{\ell}$, for all $m$, that allows the use of the addition theorem of spherical harmonics\cite{Colton:2013,Kennedy-book:2013}.
\item
Anisotropic kernels but still with the spherical harmonics as eigenfunctions, \eqref{eqn:yrkhs-kernel} and Section \ref{sec:rkhscon}.
\item
Anisotropic kernels with eigenfunctions corresponding to an arbitrary complete orthonormal sequence of functions on the sphere as described at the start of this section.
\end{enumerate}

Eigenfunctions diagonalize the integral operator associated with a kernel.  For Classes 1 and 2 above the eigenfunctions \eqref{eqn:p4} diagonalize the integral operator \eqref{eqn:integ-op} with kernel \eqref{eqn:yrkhs-kernel}. So Classes 1 and 2 above could be described as kernels diagonalized by the spherical harmonics but being isotropic and anisotropic, respectively.  Even for Class 3 spherical harmonics can be used to express the kernel but in that case the operator matrix associated with the self-adjoint integral operator is not diagonal (but still is Hermitian) and the kernel would look like\cite{Kennedy-book:2013}
\[
	\widetilde{K}(\unit{x},\unit{y}) \dfn \sum_{\ell,m} \sum_{p,q}
		\widetilde{k}_{\ell,p}^{m,q} \, Y_{\ell}^{m}(\unit{x}) \conj{Y_{p}^{q}(\unit{y})},\text{\quad with matrix\quad}
	\widetilde{k}_{\ell,p}^{m,q} \dfn \innerp{\op{L}_{\widetilde{K}}Y_{p}^{q}}{Y_{\ell}^{m}}\text{\quad satisfying\quad}
	\widetilde{k}_{\ell,p}^{m,q} = \conj{\widetilde{k}_{p,\ell}^{q,m}},
\]
but it is incompatible with Theorem \ref{thm:shrkhs}, which is limited to Classes 1 and 2.

For the remainder of this paper we restrict attention to the case of isotropic kernels or Class 1.  This leads to significant simplification and the subsequent style of analysis is almost ubiquitous in the theory of processing signals on the sphere\cite{Seon:2006,Kennedy:2011}.   In the literature, cognate with our work, the terminology ``radial basis functions'' and ``zonal kernels'', is commonly used\cite{Levesley:2005}.

\section{Closed-Form Isotropic Reproducing Kernels}

The reproducing kernel property enables simple computation of inner products of functions when expanded in functions derived from the same kernel, \eqref{eqn:innerp-ker}.  When contemplating inner product evaluation therefore it is of interest to find closed-form kernel functions rather than the general kernel expansion \eqref{eqn:yrkhs-kernel} which would involve an infinite complexity or at least possibly a high degree of computation.

\subsection{Isotropic Property}

Some simplification is necessary to obtain a class of closed-form reproducing kernels which are isotropic.  Here we use the term ``isotropic'' to refer to the property that the kernel only depends on the angle between the two arguments $\unit{x}\cdot\unit{y}$ and not the individual directions $\unit{x}$ and $\unit{y}$.

We consider the case where the eigenvalues satisfy
\begin{equation}
\label{eqn:isoeigen}
	\lambda_{\ell}^{m} = \lambda_{\ell},\quad \forall\,m\in\set{-\ell,\dotsc,\ell}
\end{equation}
and then \eqref{eqn:yrkhs-kernel} becomes
\begin{equation}
\label{eqn:yrkhs-iso-kernel}
\begin{split}
	K(\unit{x},\unit{y}) &= \sum_{\ell=0}^{\infty} \lambda_{\ell}\sum_{m=-\ell}^{\ell}
		 \, Y_{\ell}^{m}(\unit{x})   \conj{Y_{\ell}^{m}(\unit{y})} \\
		 &= k(\unit{x}\cdot\unit{y})
\end{split}
\end{equation}
where we have defined the univariate kernel function
\begin{align}
\label{eqn:yrkhs-iso-univar}
	k(z) &\dfn \frac{1}{4\pi}
		\sum_{\ell=0}^{\infty} 
				(2\ell+1)
		\lambda_{\ell}
		P_{\ell}(z),\quad -1\leq z\leq+1
\intertext{where $P_{\ell}(\cdot)$ is the Legendre polynomial of degree $\ell$ and the Legendre coefficients are}
\label{eqn:alpha}
	\alpha_{\ell} &= \frac{(2\ell+1)}{4\pi} \lambda_{\ell},
\end{align}
and we have used the addition theorem of spherical harmonics\cite{Colton:2013}.  In essence the term $(2\ell+1)$ captures the multiplicity of the $\lambda_{\ell}$ eigenvalue.  Here $\alpha_{\ell}$ is the Fourier coefficient corresponding to the use of the orthogonal Legendre polynomials.  From \eqref{eqn:alpha} it is a simple matter to relate $\alpha_{\ell}$ with $\lambda_{\ell}$.

\subsection{Hilbert-Schmidt Property for Isotropic Kernels}

We can specify the requirement \eqref{eqn:lambda-seq} to incorporate the properties that the eigenvalues are with multiplicity, real-valued and positive
\begin{equation}
\label{eqn:eigcond}
\begin{split}
	\sequ{\lambda_{\ell}^{m}= \lambda_{\ell}\colon \lambda_{\ell}^{m}\in\mathbb{R},
		\lambda_{\ell}^{m}>0}_{\ell,m}\in\ltwo  \iff 
	\sequ[\big]{ \sqrt{2\ell+1}\,\lambda_{\ell}\colon \lambda_{\ell}\in\mathbb{R},
		\lambda_{\ell}>0}_{\ell}\in\ltwo
\end{split}
\end{equation}
Considering \eqref{eqn:yrkhs-iso-univar}, this implies that the condition for the Legendre polynomial coefficients $\sequ{\alpha_{\ell}}_{\ell=0}^{\infty}$ to be finite energy differs but is easily related to the condition on the eigenvalues $\sequ{\lambda_{\ell}}_{\ell=0}^{\infty}$, \eqref{eqn:alpha}.

\subsection{Procedure for Finding Closed-Form Isotropic Kernels}

This Hilbert-Schmidt property, $\sequ{\lambda_{\ell}^{m}}_{\ell}^{m}\in\ltwo$, may at first glance appear to be difficult to work but this condition can be shown to be equivalent to the simple and natural energy condition on the (real) univariate kernel function\cite{Kennedy-book:2013}
\begin{equation}
\label{eqn:fineng}
	\int_{-1}^{+1} k^{2}(z)\,dz < \infty.
\end{equation}
So this suggests a procedure that we will follow later in trying to discover isotropic kernels that satisfy the RKHS conditions:
\begin{enumerate}[\bfseries~~Step 1.]
\item Choose a candidate closed-form univariate kernel function $k(\cdot)$ on $[-1,+1]$.
\item Confirm that $k(\cdot)$ satisfies condition \eqref{eqn:fineng}, which ensures that the eigenvalues decay to zero sufficiently quickly (Hilbert-Schmidt condition).
\item Confirm the eigenvalues are positive. This requires us to expand $k(\cdot)$ in the Legendre polynomials to determine the coefficients which through \eqref{eqn:alpha} are directly related to the eigenvalues.  That is, confirm
\begin{align*}
	\lambda_{\ell} &> 0,\quad \ell=0,1,\dotsc
\shortintertext{where\cite{Kennedy:2011}}
	\lambda_{\ell} &= 2\pi \int_{-1}^{+1} k(z) P_{\ell}(z)\,dz,\quad \ell=0,1,2,\dotsc
\end{align*}
\item
Sometimes it happens that for some candidate ${\widetilde k}(z)$ for a finite number of $\ell$ we have $\lambda_{\ell}\leq0$, call this an exception set.  In these cases, because $P_{\ell}(\cdot)$ is a special function, it is easy to modify the ${\widetilde k}(\cdot)$ so the modified univariate kernel function $k(\cdot)$ is closed-form and its eigenvalues are positive.  But if this exception set is too big then it somewhat defeats the purpose of having a closed-form kernel which preferably involves a single elementary or special function.  Some examples of doing such repairing, where the exception sets have cardinality one, are given later which highlights the usefulness of the modification procedure.
\item
Multiplying a univariate kernel function by a positive real scalar preserves the desired properties of the kernel.  A preferable scaling achieves normalization in the sense that
\[
	\int_{\untsph} k(\cos\theta)\,\sin\theta\,d\theta\,d\phi = 2\pi \int_{-1}^{+1} k(z)\,dz=1
	\iff \lambda_{0}=1,
\]
which is always possible for Hilbert-Schmidt kernels satisfying \eqref{eqn:fineng} by the Cauchy-Schwarz inequality.  This normalization is compatible with probability distributions on the 2-sphere such as the well-known \vMF distribution\cite{Seon:2006}.  But to be a probability distribution requires the additional condition $k(z)\geq0$ for all $z\in[-1,+1]$.
\item
Synthesize the kernel from the univariate kernel functions $k(\cdot)$ by writing it in the $K(\unit{x},\unit{y})=k(\unit{x}\cdot\unit{y})$, as given in \eqref{eqn:yrkhs-iso-kernel}.
\end{enumerate}

In fact the goal of the isotropic kernel construction above can be written in a simple self-contained way that makes no specific demand on understanding the underlying theory:

\vspace{3mm}
\begin{theorem}
\label{thm:uniker}
 A univariate kernel function $k(\cdot)$ defined on $[-1,+1]$ satisfying the admissibility conditions:
\[
	\int_{-1}^{+1} k^{2}(z)\,dz < \infty \text{\quad and\quad}
	\int_{-1}^{+1} k(z) P_{\ell}(z)\,dz > 0,\quad \forall\,\ell=0,1,\dotsc
\]
leads to an isotropic reproducing kernel, $K(\unit{x},\unit{y})=k(\unit{x}\cdot\unit{y})$, and conversely.
\end{theorem}
\vspace{3mm}

A fast-track alternative to the above, which proves to be fruitful, is to harvest the literature for Legendre series expansions of elementary closed-form functions where all but a finite number (preferably only a few) of the Legendre-Fourier coefficients are of the same sign and non-zero.  Modifying such functions in an obvious way leads to admissible univariate kernel functions $k(\cdot)$ and is illustrated later.

\section{Explicit Constructions}

Any sufficiently regular function $k(\cdot)$ defined on domain $[-1,+1]$ can be expanded in terms of Legendre polynomials.  For our purposes we require the Legendre-Fourier coefficients to be positive and the corresponding eigenvalues $\sequ{\lambda_{\ell}}$ to decay sufficiently quickly.  But, more critically we need the kernel to manifest itself in terms of a single (or few) elementary closed-form functions.

\begin{figure}[tbp]
\centering
	\includegraphics[scale=1.0]{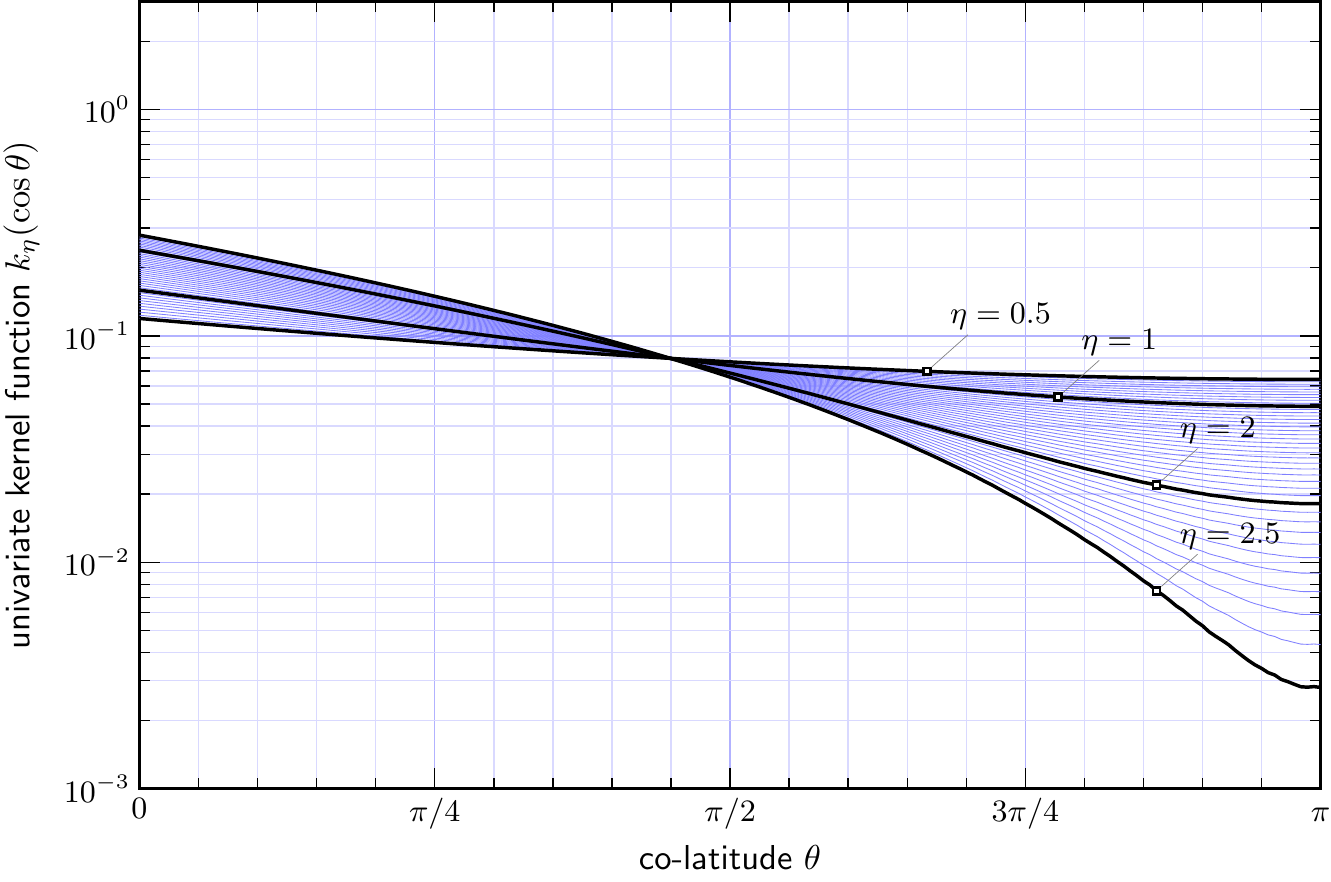}\\
	\caption{Cui and Freden univariate kernel functions, $k_{\eta}(\cos\theta)\dfn(1+\eta\,\widetilde{k}(z))/(4\pi)$, for $\eta=0.5,1,2,2.5$  (intermediate values increment by $0.05$) plotted against co-latitude $\theta$.}\label{fig:chi}
\end{figure}

\subsection{Cui and Freden Kernel}

Motivated by finding the integral operator corresponding to a single-layer potential, the identity
\begin{equation}
\label{eqn:chi}
	\widetilde{k}(z) \dfn 1 - 2\log\parn[\Big]{1+\sqrt{\frac{1-z}{2}}}
		= \sum_{\ell=1}^{\infty} \frac{P_{\ell}(z)}{\ell(\ell+1)},\quad -1\leq z\leq+1
\end{equation}
has been used to construct a RKHS kernel on the 2-sphere\cite{Cui:1997}.  Evidently because $\ell(\ell+1)>0$ for all $\ell=1,2,\dotsc$ (the exception set is for $\ell=0$ where the coefficient is zero), then it is easily modified to form an admissible univariate kernel function.

From \eqref{eqn:chi}, adding a $P_{0}(z)$ term and scaling by $1/(4\pi)$, we can use \eqref{eqn:yrkhs-iso-univar} to infer that the eigenvalues
\begin{equation}
\label{eqn:chi-eigenv}
	\lambda_{\ell} =
	\begin{cases}
		1 &\ell=0 \\
		\displaystyle\frac{1}{\ell(\ell+1)(2\ell+1)} &\ell=1,2,\dotsc
	\end{cases}
\end{equation}
directly satisfy \eqref{eqn:eigcond} but, of course, \eqref{eqn:fineng} provides an easier demonstration.  The associated univariate kernel function is
\[
	k(z) = \frac{1}{4\pi}\parn[\Big]{1+\widetilde{k}(z)} = \frac{1}{2\pi} -
		\frac{1}{2\pi}\log\parn[\Big]{1+\sqrt{\frac{1-z}{2}}},
\]
which satisfies the normalization (equivalent to the condition $\lambda_{0}=1$)
\[
	2\pi \int_{-1}^{+1} k(z)\,dz=1.
\]
Then the associate kernel, by \eqref{eqn:yrkhs-iso-kernel}, is\cite{Cui:1997}
\begin{equation}
\label{eqn:chi-kernel}
	K(\unit{x},\unit{y}) \dfn \frac{1}{2\pi}
		- \frac{1}{2\pi}\log\parn[\Big]{1+\sqrt{\frac{1-(\unit{x}\cdot\unit{y})}{2}}}
\end{equation}
is a reproducing kernel.

It is possible to create a family of admissible kernels, which includes \eqref{eqn:chi-kernel} as a special case, by positive weighting of the $\widetilde{k}(z)$ portion with some positive parameter $\eta$, leading to the family $k_{\eta}(\cos\theta)\dfn(1+\eta\,\widetilde{k}(z))/(4\pi)$.  (This type of modification is illustrated in the Lebedev kernel that we reveal in the next subsection.)  We plot such a univariate kernel function family for $\eta=0.5,1,2,2.5$ (intermediate values, which increment by $0.05$, are also shown), in \figref{fig:chi}.

Finally, as guaranteed by the developed theory, $Y_{\ell}^{m}(\unit{x})$ is an eigenfunction of the integral equation with the above kernel with eigenvalue $\lambda_{\ell}>0$, for all orders $-\ell\leq m\leq\ell$, given in \eqref{eqn:chi-eigenv}.

\begin{figure}[tbp]
\centering
	\includegraphics[scale=1.0]{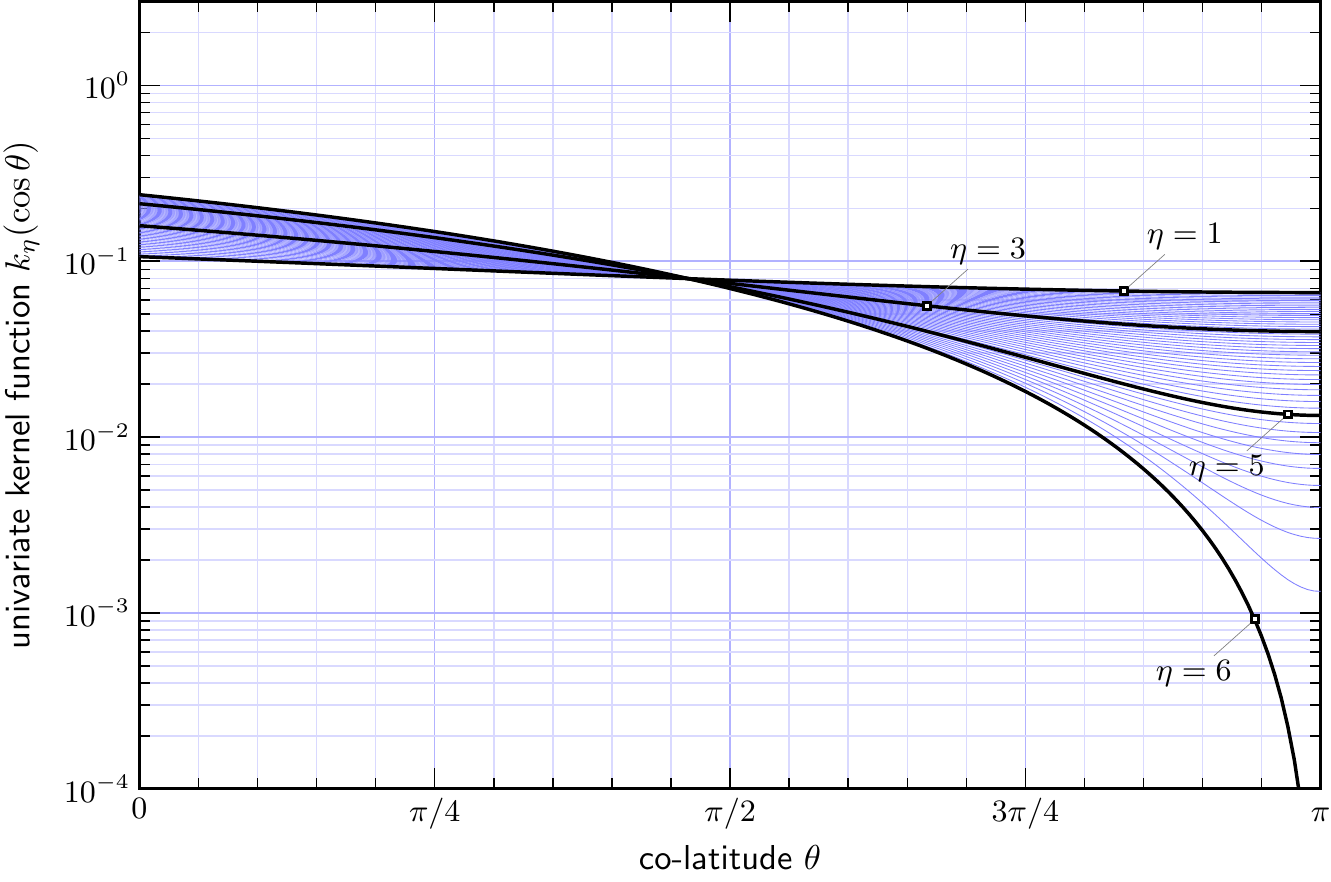}\\
	\caption{Lebedev univariate kernel functions, $k_{\eta}(\cos\theta)$, for $\eta=1,2,4,6$  (intermediate values increment by $0.1$) plotted against $z=\cos\theta$ where $\theta$ is the co-latitude.}\label{fig:lebedev}
\end{figure}

\subsection{Lebedev Kernel}

The Cui and Freden kernel\cite{Cui:1997} is from the literature but we can find new kernels via similar means.  The next identity is based on the Legendre polynomial expansion\cite{Lebedev:1972,Gradshteyn:2007}
\[
	\sqrt{\frac{1-z}{2}} = \frac{2}{3} P_{0}(z) -
		2 \sum_{\ell=1}^{\infty} \frac{P_{\ell}(z)}{(2\ell-1)(2\ell+3)},\quad -1\leq z\leq+1.
\]
Apart from the first term, where the coefficient for $P_{0}(z)$ is positive, all coefficients are of the same sign (negative).  Define, noting that $P_{0}(z)=1$,
\[
	\widetilde{k}(z) \dfn
		\frac{1}{3} - \frac{1}{2} \sqrt{\frac{1-z}{2}} =
		\sum_{\ell=1}^{\infty} \frac{P_{\ell}(z)}{(2\ell-1)(2\ell+3)},\quad -1\leq z\leq+1,
\]
which has Legendre Fourier coefficients and eigenvalues, according to \eqref{eqn:alpha},
\[
	\widetilde{\alpha}_{\ell} =
	\begin{cases}
		\displaystyle 0 &\ell=0 \\
		\displaystyle\frac{1}{(2\ell-1)(2\ell+3)} &\ell=1,2,\dotsc
	\end{cases}, \text{\quad and\quad}
	\widetilde{\lambda}_{\ell} =
	\begin{cases}
		\displaystyle 0 &\ell=0 \\
		\displaystyle\frac{4\pi}{(4\ell^{2}-1)(2\ell+3)} &\ell=1,2,\dotsc
	\end{cases}
\]
Because $\widetilde{\alpha}_{0}=0$ is the coefficient of $P_{0}(z)=1$, which is orthogonal to all other $P_{\ell}(z)$ for $\ell=1,2,\dotsc$, then
\(
	2\pi \int_{-1}^{+1} \widetilde{k}(z)\,dz=0,
\)
and evidently normalization to unity is only achieved by introducing a $P_{0}(z)$ term corresponding to $\lambda_{0}=1$.  So define our univariate kernel function, parameterized by $\eta>0$, as
\begin{align*}
	k_{\eta}(z)
		&\dfn \frac{1}{4\pi}\parn[\Big]{1+\eta\,{\widetilde{k}(z)}} \\
		&= \parn[\Big]{\frac{1}{4\pi}+\frac{\eta}{12\pi}}
			- \frac{\eta}{8\pi}\sqrt{\frac{1-z}{2}},
\end{align*}
so as to have eigenvalues
\[
	\lambda_{\ell}(\eta) =
	\begin{cases}
		\displaystyle 1 &\ell=0 \\
		\displaystyle\frac{\eta}{(4\ell^{2}-1)(2\ell+3)} &\ell=1,2,\dotsc
	\end{cases}
\]
and to satisfy (equivalent to the condition $\lambda_{0}(\eta)=1$)
\[
	2\pi \int_{-1}^{+1} k_{\eta}(z)\,dz=1.
\]
We also note that $k_{\eta}(z)\geq0$ for all $z$ whenever $0<\eta\leq6$, so for this range $k_{\eta}(z)$ can serve as a probability distribution.  We plot $k_{\eta}(\cos\theta)$, for $\eta=1,2,4,6$ (intermediate values, which increment by $0.1$, are also shown), in \figref{fig:lebedev} where evidently varying $\eta$ gives limited control of the shape.

Finally the kernel given by
\begin{equation}
\label{eqn:leb-kernel}
	K_{\eta}(\unit{x},\unit{y}) \dfn
		\parn[\Big]{\frac{1}{4\pi}+\frac{\eta}{12\pi}}
			- \frac{\eta}{8\pi}\sqrt{\frac{1-(\unit{ x}\cdot\unit{y})}{2}}
\end{equation}
is a reproducing kernel for any $\eta>0$.  In consideration of the computations required for the inner product in the RKHS, \eqref{eqn:innerp-ker}, then the \eqref{eqn:leb-kernel} kernel is more elementary than \eqref{eqn:chi-kernel}, which requires $\log(\cdot)$ evaluations.

\begin{figure}[tbp]
\centering
	\includegraphics[scale=1.0]{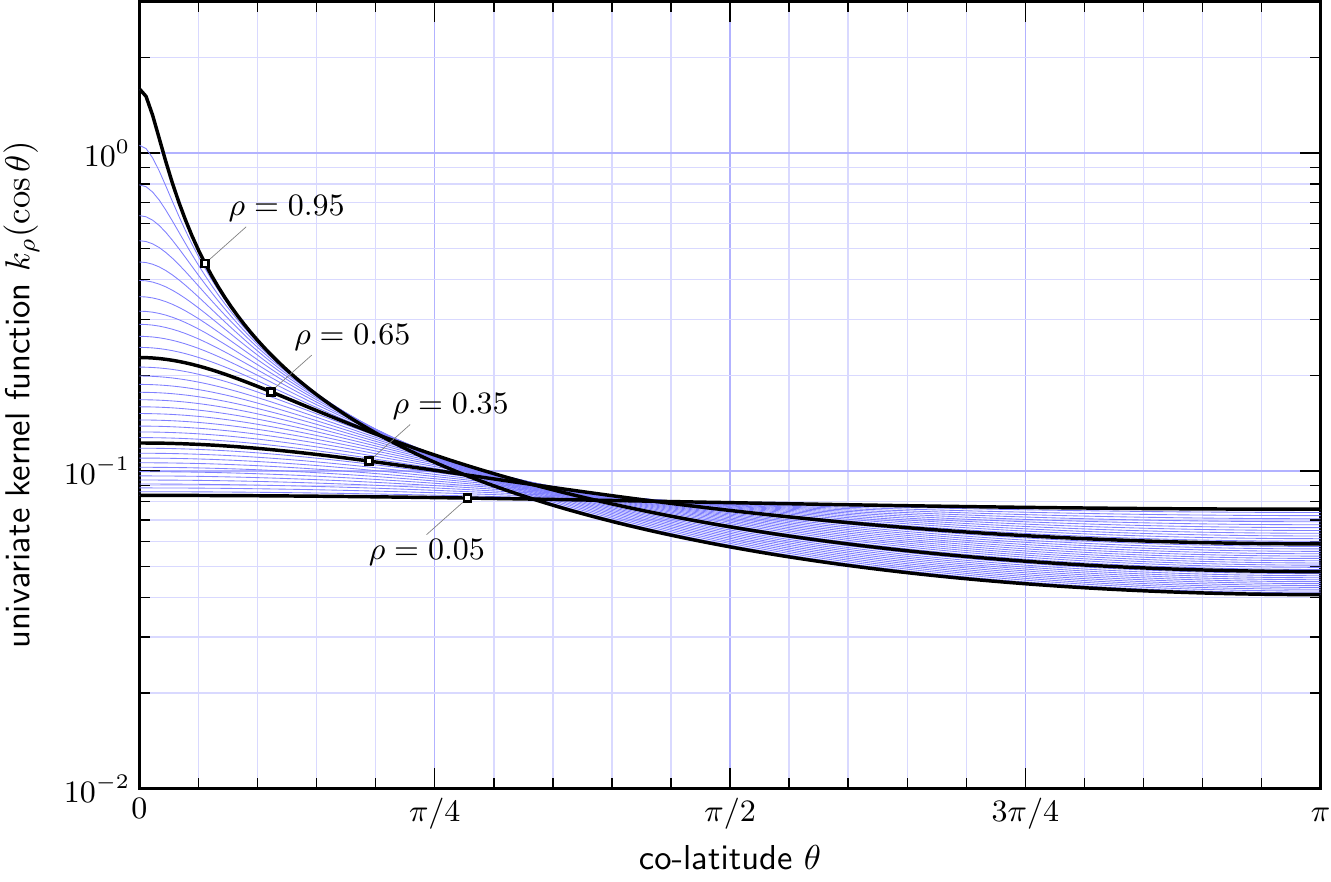}\\
	\caption{Legendre univariate kernel functions for $\rho=0.05,0.35,0.65,0.95$ (intermediate values increment by $0.025$) plotted against co-latitude $\theta$.}\label{fig:leg-gen}
\end{figure}

\subsection{Legendre Generating Function Kernels}

The previous kernels exhibit polynomial decay, in fact cubic decay, of the eigenvalues (which is really a quadratic decay when one takes into account the multiplicity of the eigenvalues).  The decay rate corresponds to the degree of smoothness of the functions in the RKHS where, of course, a faster decay leads to a greater smoothness (or, very loosely, what has been described as a smaller space).  Decay rates are known to relate to the function class $C^{k}$ which correspond to functions where derivatives up to $k$ exists and are continuous\cite{Strang:2007,Kennedy-book:2013}.  This means the functions involved may be discontinuous once a sufficient number of derivatives, greater in number than $k$, are taken (but these can hardly be thought of as non-smooth functions).  However, one can seek RKHS function spaces where all derivatives exists and this requires a stronger sense of convergence than polynomial decay. This partially motivates the search for eigenvalue decay rates which are exponential.

A second limitation with the polynomial decay kernels, \eqref{eqn:chi-kernel} and \eqref{eqn:leb-kernel}, is they are not very flexible in their shape or more specifically their spread. (One can adjust a constant factor corresponding to $\lambda_{0}>0$ which is not particularly helpful.)  Such RKHS kernels are often thought of as interpolating or approximating functions as in \eqref{eqn:fexpan}.  This is, the $\unit{y}_{p}\in\untsph$, for $p=1,2,\dotsc,P$, should relate to the spread of the kernel.  This provides a second motivation to look for closed-form kernels which are parameterized by some continuous parameter that adjusts the spread of the kernel.  Of course for every instance of the spread parameter we have strictly speaking a different RKHS. 

In this case, to devise a closed-form kernel, we propose to use the generating function for the Legendre polynomials, which is given by
\begin{equation}
\label{eqn:leggen}
	\widetilde{k}_{\rho}(z)
		\dfn \frac{1}{\sqrt{1-2\,z\,\rho+\rho^{2}}}
		= \sum_{\ell=0}^{\infty} \rho^{\ell}\,P_{\ell}(z),\quad -1\leq z\leq+1.
\end{equation}
The key observation is that the coefficients, $\rho^{\ell}$, are all positive whenever parameter $\rho>0$.  We scale this to define
\[
	k_{\rho}(z) \dfn \frac{\widetilde{k}_{\rho}(z)}{4\pi} = \frac{1}{4\pi\sqrt{1-2\,z\,\rho+\rho^{2}}},
	\text{\quad such that\quad} 2\pi\int_{-1}^{+1} k_{\rho}(z)\,dz=1 \iff \lambda_{0}=1.
\]
Then it is a simple matter to identify the Legendre coefficients of $k_{\rho}(z)$ and the eigenvalues using \eqref{eqn:alpha}
\[
	\alpha_{\ell} = \frac{\rho^{\ell}}{4\pi} \text{\quad and\quad}
	\lambda_{\ell} = \frac{\rho^{\ell}}{2\ell+1}, \quad \ell=0,1,2,\dotsc.
\]
By taking $0<\rho<1$ (Hilbert-Schmidt condition) then $\lambda_{\ell}>0$ (all eigenvalues are positive and decay exponentially).  Then
\begin{equation}
\label{eqn:gen-kernel}
	K_{\rho}(\unit{x},\unit{y}) \dfn \frac{1}{4\pi\sqrt{1-2\,(\unit{x}\cdot\unit{y})\,\rho+\rho^{2}}},\quad 0<\rho<1
\end{equation}
is a reproducing kernel for each suitable $\rho$.  \figref{fig:leg-gen} shows the univariate kernel functions for $\rho=0.05,0.35,0.65,0.95$ (intermediate values, which increment by $0.025$, are also shown).


From the generating function we can construct further closed-form RKHS kernels through simple manipulations. For example, taking first derivative
\[
	\frac{\partial}{\partial \rho} \sum_{\ell=0}^{\infty} \rho^{\ell}\,P_{\ell}(z)
		= \frac{\partial}{\partial \rho} \frac{1}{\sqrt{1-2\,z\,\rho+\rho^{2}}}
\]
leads to
\[
	\sum_{\ell=1}^{\infty} \ell\, \rho^{\ell-1}\,P_{\ell}(z)
		= \frac{(\rho-z)}{(1-2\,\rho\,z+\rho^{2})^{3/2}}
\]
where evidently the Legendre coefficients are positive $\ell\rho^{\ell-1}>0$ for all $\ell=1,2,\dots$ and $\ell=0$ is an exception set.  As for some of the earlier kernels we can add the term $\lambda_{0}P_{0}(z)=\lambda_{0}>0$ to this expression to make it an admissible RKHS univariate kernel function.
This procedure can be generalized by applying a general $P$-order differential operator with respect to $\rho$ such as
\[
	\parn[\big]{\sum_{p=0}^{P} \chi_{p}\frac{\partial^{p}}{\partial\rho^{p}}}
		\sum_{\ell=0}^{\infty} \rho^{\ell}\,P_{\ell}(z),
\]
where the differential operator coefficients $\chi_{p}$ can always be chosen such that the overall coefficients of $P_{\ell}(z)$ in the expression are positive.

\begin{figure}[tbp]
\centering
	\includegraphics[scale=1.0]{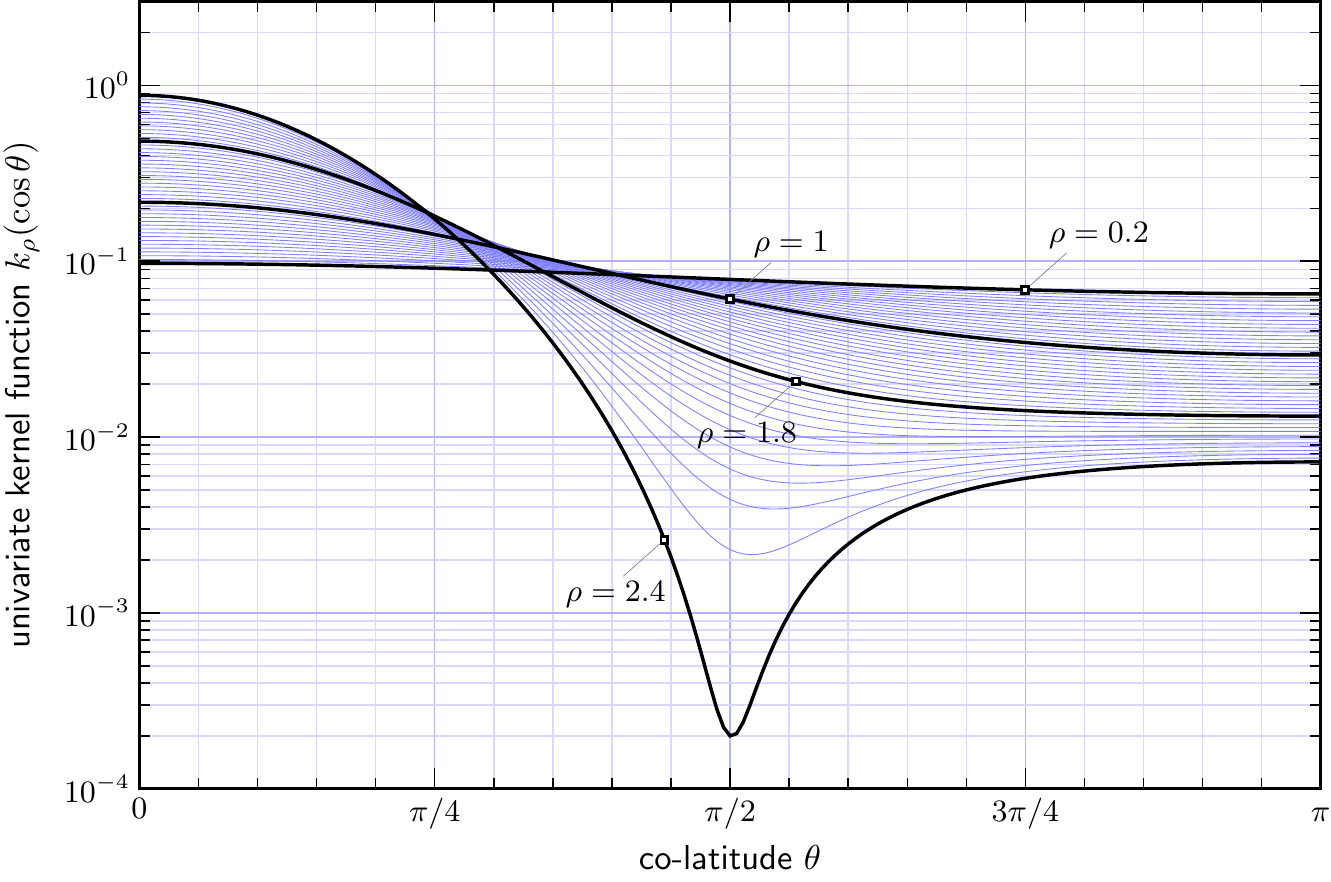}\\
	\caption{Alternative Legendre generating function univariate kernel functions for $\rho=0.2,1,1.8,2.4$ (intermediate values increment by $0.05$) plotted against co-latitude $\theta$.}\label{fig:altgen}
\end{figure}

\subsection{Alternative Legendre Generating Function Based Kernels}

Another generating function for the Legendre polynomials is given by\cite{Koepf:1998}
\begin{equation}
\label{eqn:leggenalt}
	\widetilde{k}_{\rho}(z)
		\dfn e^{\rho z} J_{0}\parn[\big]{\rho\sqrt{1-z^{2}}}
		= \sum_{\ell=0}^{\infty} \frac{1}{\ell\,!}\rho^{\ell}\,P_{\ell}(z),\quad -1\leq z\leq+1,
\end{equation}
where for $\rho>0$ all Legendre coefficients are positive.  Then $k_{\rho}(z)=\widetilde{k}_{\rho}(z)/4\pi$ and
\[
	K_{\rho}(\unit{x},\unit{y}) \dfn \frac{e^{\rho\, \unit{x}\cdot\unit{y}}}{4\pi}
		J_{0}\parn[\big]{\rho\sqrt{1-(\unit{x}\cdot\unit{y})^{2}}}
\]
is a reproducing kernel with positive eigenvalues
\[
	\lambda_{\ell}(\rho) = \frac{\rho^{\ell}}{(2\ell+1)\,\ell!},\quad \ell=0,1,2,\dotsc,
		\text{\quad noting\quad}
	\lambda_{\ell}(\rho) \sim \frac{1}{2\sqrt{2\pi}(\rho\,e)^{3/2}} \parn[\Big]{\frac{\rho\,e}{\ell}}^{\ell+3/2}
		\text{ as }\ell\tendsto\infty,
\]
where clearly for all parameter values $\rho>0$ the Hilbert-Schmidt condition, \eqref{eqn:eigcond}, is satisfied.  In \figref{fig:altgen} we plot
\[
	k_{\rho}(\cos\theta) = \frac{e^{\rho \cos\theta}}{4\pi} J_{0}\parn[\big]{\rho\sin\theta}
\]
for $\rho=0.2,1,1.8,2.4$ (intermediate values, which increment by $0.05$, are also shown).

\begin{figure}[tbp]
\centering
	\includegraphics[scale=1.0]{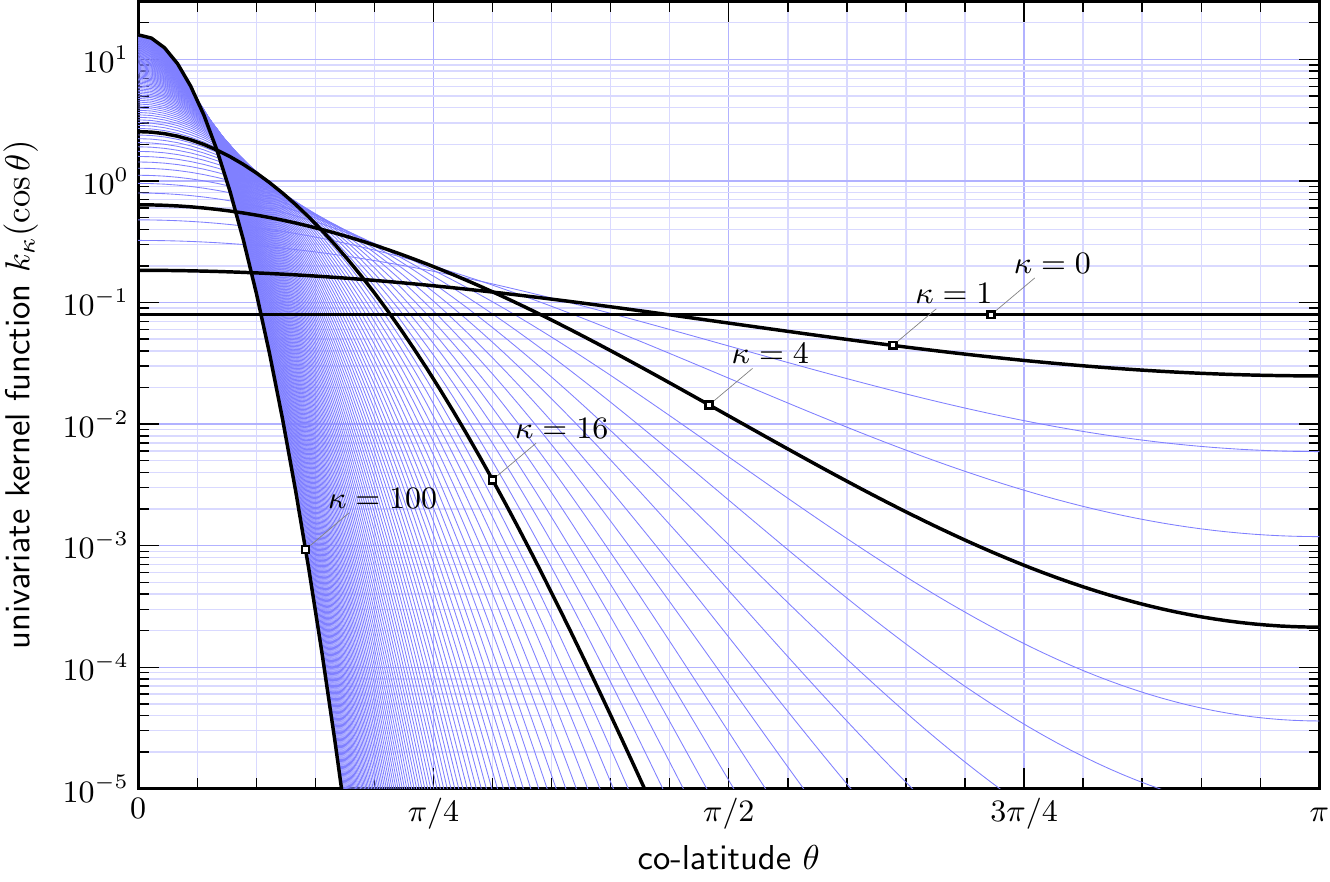}\\
	\caption{The von Mises-Fisher univariate kernel functions for $\kappa=0,1,4,16,100$ (intermediate values increment by $1$) plotted against co-latitude $\theta$.}\label{fig:vonMises}
\end{figure}

\subsection{von Mises-Fisher Kernels}

The final closed-form kernel is motivated by the desire to have a kernel with adjustable spread and be an analogy of gaussian kernels defined on the real line.  The \vMF distribution\cite{Seon:2006} is given by
\[
	k_{\kappa}(z) \dfn \frac{\kappa\exp(\kappa\,z)}{4\pi\sinh\kappa}
	\text{\quad such that\quad} 2\pi\int_{-1}^{+1} k_{\kappa}(z)\,dz=1,
\]
where the concentration parameter is given by $\kappa\geq0$ and is an admissible RKHS univariate kernel function.  The calculation for the Legendre coefficients and eigenvalues can be found in the literature.  A recursive method generates the eigenvalues in order as follows\cite{Seon:2006}
\begin{align*}
	\lambda_{0}(\kappa) = 1,~ \lambda_{1}(\kappa) = \coth\kappa - \frac{1}{\kappa},~
	\lambda_{2}(\kappa) = \frac{\kappa^{2}-3\kappa\coth\kappa+3}{\kappa^{2}},~\dotsc,
\end{align*}
under the recursion
\[
	\lambda_{\ell+1}(\kappa) = \lambda_{\ell-1}(\kappa) - \frac{(2\ell+1)}{\kappa}\lambda_{\ell}(\kappa),
		\quad\ell=1,2,\ldots,
\]
from which it can be established\cite{Kennedy-book:2013} $\lambda_{\ell}(\kappa)>0$, for all $\ell=0,1,2,\dotsc$, and for all $\kappa\geq0$, using a generating function\cite{Seon:2006}.  These eigenvalues can also be expressed in closed form as\cite{Abramowitz:1970,Mammasis:2010}
\[
	\lambda_{\ell}(\kappa) = \frac{I_{\ell+1/2}(\kappa)}{I_{1/2}(\kappa)}
		= \sqrt{\frac{\pi\kappa}{2}} \frac{I_{\ell+1/2}(\kappa)}{\sinh\kappa}
\]
where $I_{\ell+1/2}(\cdot)$ is a half-integer-order modified Bessel function of the first kind.

So we have a family of reproducing kernels
\begin{equation}
\label{eqn:von-kernel}
	K_{\kappa}(\unit{x},\unit{y}) \dfn 
		\frac{\kappa\exp(\kappa\,\unit{x}\cdot\unit{y})}{4\pi\sinh\kappa},\quad \kappa\geq0,
\end{equation}
and each value of $\kappa$ yields a different spread kernel, as illustrated in \figref{fig:vonMises} for $\kappa=0,1,4,16,100$ (intermediate values, which increment by $1$, are also shown), and a different RKHS.

\section{Conclusions}

We have provide an accessible framework where closed-form kernels can be constructed and interpreted so as to define various reproducing kernel Hilbert spaces on the 2-sphere.  This leads to a classification of such kernels into three classes: kernels where the associated integral operator are diagonalized by the spherical harmonics, which can either be isotropic or anisotropic, and more general kernels.  For the isotropic case, conditions on a univariate kernel function were given and this led to a procedure to help construct closed-form isotropic kernels.  Five different kernels or kernel families were constructed adding to known kernels from the literature.


\acknowledgments     
 
This work was partially supported under the Australian Research Council's Discovery Projects funding scheme (Project No.~DP1094350).  Jason D. McEwen is supported in part by a Newton International Fellowship from the Royal Society and the British Academy. 


\bibliography{s2-kernels}   

\begin{thebibliography}{10}

\bibitem{Lebedev:1972}
Lebedev, N.~N.,  [{\em Special Functions and Their
  Applications}{\nolinebreak\hspace{0.1em}]}, Dover Publications, New York, NY
  (1972).

\bibitem{Kennedy-book:2013}
Kennedy, R.~A. and Sadeghi, P.,  [{\em Hilbert Space Methods in Signal
  Processing}{\nolinebreak\hspace{0.1em}]}, Cambridge University Press,
  Cambridge, UK (Mar. 2013).

\bibitem{Aronszajn:1950}
Aronszajn, N., ``Theory of reproducing kernels,'' {\em Trans. Amer. Math.
  Soc.}~{\bf 68}(3),  337--404 (1950).

\bibitem{Cucker:2002}
Cucker, F. and Smale, S., ``On the mathematical foundations of learning,'' {\em
  Bull. Am. Math. Soc., New Ser.}~{\bf 39}(1),  1--49 (2002).

\bibitem{Colton:2013}
Colton, D. and Kress, R.,  [{\em Inverse Acoustic and Electromagnetic
  Scattering Theory}{\nolinebreak\hspace{0.1em}]}, Springer, New York, NY,
  3rd~ed. (2013).

\bibitem{Seon:2006}
Seon, K.-I., ``Smoothing of all-sky survey map with {F}isher-von {M}ises
  function,'' {\em J. Korean Phys. Soc.}~{\bf 48},  331--334 (Mar. 2006).

\bibitem{Kennedy:2011}
Kennedy, R.~A., Lamahewa, T.~A., and Wei, L., ``On azimuthally symmetric
  2-sphere convolution,'' {\em Digital Signal Process.}~{\bf 5},  660--666
  (Sept. 2011).

\bibitem{Levesley:2005}
Levesley, J. and Sun, X., ``Approximation in rough native spaces by shifts of
  smooth kernels on spheres,'' {\em J. Approx. Theory}~{\bf 133}(2),  269--283
  (2005).

\bibitem{Cui:1997}
Cui, J. and Freeden, W., ``Equidistribution on the sphere,'' {\em SIAM J. Sci.
  Comput.}~{\bf 18},  595--609 (Mar. 1997).

\bibitem{Gradshteyn:2007}
Gradshteyn, I. and Ryzhik, I.,  [{\em Table of Integrals, Series, and
  Products}{\nolinebreak\hspace{0.1em}]}, Academic Press, 7~ed. (2007).

\bibitem{Strang:2007}
Strang, G.,  [{\em Computational Science and
  Engineering}{\nolinebreak\hspace{0.1em}]}, Wellesley-Cambridge Press,
  Wellesley, MA (2007).

\bibitem{Koepf:1998}
Koepf, W.,  [{\em Hypergeometric summation: An Algorithmic Approach to
  Summation and Special Function Identities}{\nolinebreak\hspace{0.1em}]},
  Vieweg, Braunschweig (1998).

\bibitem{Abramowitz:1970}
Abramowitz, M. and Stegun, I.,  [{\em Handbook of Mathematical
  Functions}{\nolinebreak\hspace{0.1em}]}, Dover Publishing Inc., New York, NY
  (1970).

\bibitem{Mammasis:2010}
Mammasis, K. and Stewart, R.~W., ``Spherical statistics and spatial correlation
  for multielement antenna systems,'' {\em EURASIP J. Wirel. Commun.}~{\bf
  2010} (Dec. 2010).

\end{thebibliography}
\bibliographystyle{spiebib}   

\end{document}